# Comparisons of Fuzzy MRAS and PID Controllers for EMS Maglev Train

Mustefa Jibril[1]    Eliyas Alemayehu Tadese[2]
1. Msc, Department of Electrical & Computer Engineering, Dire Dawa Institute of Technology, Dire Dawa, Ethiopia
2. Msc, Faculty of Electrical & Computer Engineering, Jimma Institute of Technology, Jimma, Ethiopia

**Abstract**
In this paper, a Magnetic Levitation (MAGLEV) train is designed with a first degree of freedom electromagnet-based totally system that permits to levitate vertically up and down. Fuzzy logic, PID and MRAS controllers are used to improve the Magnetic Levitation train passenger comfort and road handling. A matlab Simulink model is used to compare the performance of the three controllers using step input signals. The stability of the Magnetic Levitation train is analyzed using root locus technique. Controller output response for different time period and change of air gap with different time period is analyzed for the three controllers. Finally the comparative simulation and experimental results demonstrate the effectiveness of the presented fuzzy logic controller.

## 1 Introduction

Magnetic levitation is the process of levitating an item via exploiting magnetic fields. If the magnetic force of enchantment is used, it is recognized as magnetic suspension. If magnetic repulsion is used, its miles referred to as magnetic levitation.

Magnetically Levitated (Maglev) trains fluctuate from traditional trains in that they are levitated, guided and propelled alongside a guide manner by means of a converting magnetic field as opposed to through steam, diesel or electric powered engine.

The magnetic levitation machine is a difficult nonlinear mechatronic machine in which an electromagnetic pressure is needed to suspend an item in the air and it calls for an excessive-overall performance controller to control the modern via the superconducting magnets.

This research is aimed at developing methods of improving efficiency in transportation. Additional applied technologies that may have uses in other applications, from inter-satellite communications, to magnetic field probes.

The two main types of maglev Technology are:
- Electromagnetic suspension (EMS):
  Makes use of attractive pressure machine to levitate. Which is a German generation.
- Electrodynamic suspension (EDS): uses repulsive force device to levitate. Which is a Japan generation.

## 2 Mathematical Models
### 2.1 Maglev train system mathematical model

The electromagnetic pressure f (i, z), acts on the train, which can be expressed as the subsequent dynamic system in upward course consistent with Newton's law:

$$m\frac{d^2 z(t)}{dt^2} = mg - f(i,z)$$

Where m is the mass of the automobile and g is the gravitational steady.
The electromagnetic force
$$f(i,z) = -\frac{i^2(t)}{2}\frac{dL(z)}{dz}\Big|_{i=constant\ for\ linear\ system}$$
The voltage-current relationship for the coil is given by

$$V(t) = Ri(t) + L(z)\frac{di(t)}{dt}$$

The displacement of the train is measured by using the sensor image-detector that is the output and can be formulated as:

$$Y = V_z(z) = \beta z$$

Where
$\beta$ is the sensor gain
The basic transfer function among the coil input voltage V(s) and the sensor output voltage Vz(s) is given as

$$G(s) = \frac{V_z(s)}{V(s)} = -\frac{K_I \beta}{(R + sL_1)(ms^2 - K_z)}$$

## 3 The Proposed Controller Design
There are two approaches of control system design.

### 3.1 Outward approach:
Is a manipulate design approach that begins from interior to outward i.e. First the open loop transfer function is shaped by controlling it poles and zeros, adding right control design to the system, so that stable normal transfer function might be achieved.

### 3.2 Inward approach:
Is the reverse of the outward technique i.e. First a preferred closed loop transfer function is designed, and then remedy for required controller.

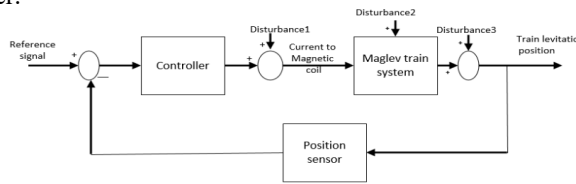

Fig 1. Block Diagram of Closed Loop Maglev Train Control System

### 3.3 Stability of maglev train system
The maglev train system model has been represented by a transfer function G(s).

$$G(s) = \frac{Y(s)}{U(s)} = \frac{-280}{(s + 29)(s + 56)(s - 56)}$$

The system has zeros at s = -29 and have poles at s = −56, and s = 56. From this, the system has a pole on the right hand side of the s-plane and this is not stable.

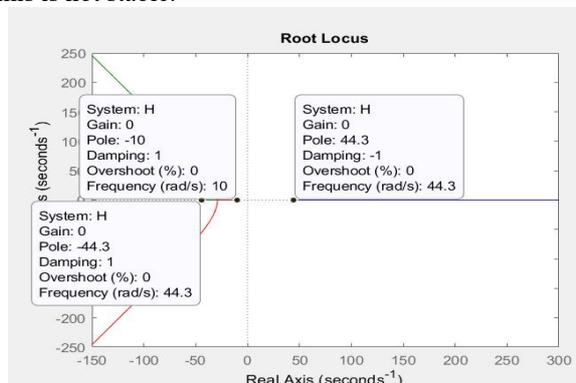

Fig 2. Root locus stability of maglev train system

### 3.4 Fuzzy Controller
The fuzzy logic control block diagram is shown in Figure 3 below.

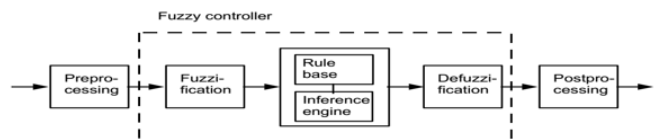

Fig 3. Block diagram of fuzzy logic Controller

The Simulink model of the fuzzy logic controller is shown in Figure 4 below.

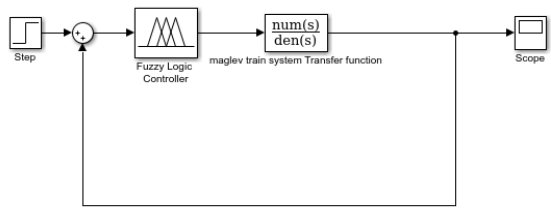

Fig 4. Simulink model of the fuzzy logic controller

### 3.4.1 Input and Output of fuzzy controller
The error and change of error input and the output of the fuzzy logic controller is shown in Figure 5, Figure 6 and Figure 7 respectively.

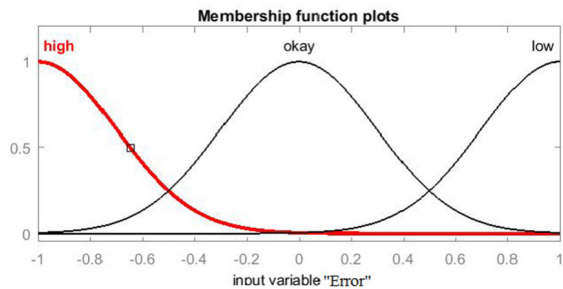

Fig 5. Error input

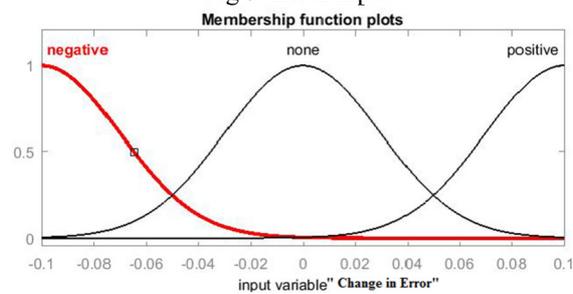

Fig 6. Change in error input

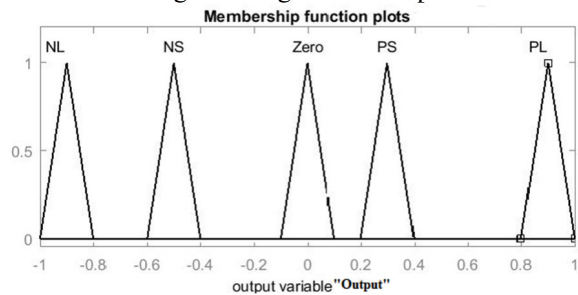

Fig 7. Output

The rule base of the fuzzy controller is shown in Table 1 below.

| No | Rules |
|---|---|
| 1 | If (error is okay) then (output is Zero) (1) |
| 2 | If (error is low) then (output is PL) (1) |
| 3 | If (error is high) then (output is NL) (1) |
| 4 | If (error is okay) and (change of error is positive) then (output is NS) (1) |

## 3.5 MRAS
**Modified MIT Rule**
Normalization can be used to protect against dependence on the signal amplitudes
Consider the first-order system

$$y(s) = \frac{u}{s + 0.65}$$

$$\frac{dy}{dt} = -ay + bu$$

$$\frac{dy}{dt} = -0.65y + u$$

$$a = 0.65$$

$$b = 1$$

The desired closed-loop system is

$$ym(s) = \frac{5Uc}{s+2}$$

$$\frac{dym}{dt} = -am\,ym + bm\,uc$$

$$\frac{dym}{dt} = -2ym + 5uc$$

$$am = 2$$

$$bm = 5$$

The controller is then

$$u(t) = to\,Uc(t) - so\,y(t)$$

$$U = to Uc - so Y = Y\frac{s+a}{b}$$

$$\frac{Y}{Uc} = \frac{to}{\frac{s+a}{b}+so} = \frac{b\,to}{s+a+bso}$$

$$bm = b\,to$$

$$am = a + bs0$$

$$e = Y - Ym = \frac{b\,to}{s+a+bso}Uc = \frac{b}{s+am}Uc$$

$$\frac{\partial e}{\partial to} = \frac{b}{s+a+bso}Uc = \frac{b}{s+am}Uc$$

$$\approx Gm(s)Uc = ym$$

$$\frac{\partial e}{\partial so} = \frac{-b^2 to}{(s+a+bso)^2}Uc = \frac{-b}{s+a+bso}Y = \frac{-b}{s+am}Y$$

$$\approx -Gm(s)Y$$

With

$$\varphi = \frac{\partial e}{\partial \theta}$$

The MIT rule can be written as

$$\frac{d\theta}{dt} = -\gamma \varphi e$$

The normalized MIT rule is then

$$\frac{d\theta}{dt} = \frac{-\gamma \varphi e}{\alpha^2 + \varphi^T \varphi}$$

$$\frac{dto}{dt} = -\gamma_1 \frac{y_m e}{\alpha_1 + y_m^2}$$

$$\frac{dso}{dt} = \gamma_2 \frac{Gm(s)ye}{\alpha_2 + (Gm(s)y)^2}$$

The Simulink model of the MRAS controller is shown in the Figure 8 bellow.

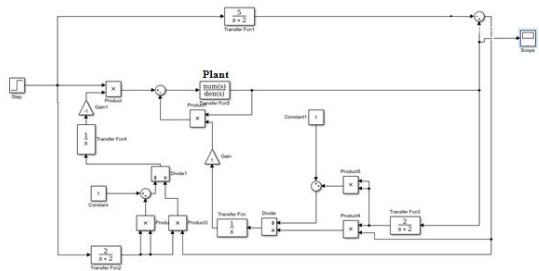

Fig 8 Simulink model of the Mras controller

**3.6 PID**
The PID (Proportional-Integral-Differential) regulator manipulate depending on the proportional, essential and differential of the deviation
General equation of PID:
$$Output = K_p e(t) + K_I \int e(t)dt + K_D \frac{d}{dt}e(t)$$
Where: $e = Setpoint - Input$

**3.6.1 PID Tuning**
The ZNFD approach may be tough to perform because it is intricate to modify the advantage till the close-loop system oscillates. A little beyond that outcomes causes instability.
   The reaction of automatic tuning is exceptionally exact whilst in comparison to the reaction of Ziegler Nichols. So, automatic tuning is used in matlab is used to stabilize the system. Based at the parameters discovered from automobile tuning, attempt to error method is used until higher result is achieved.

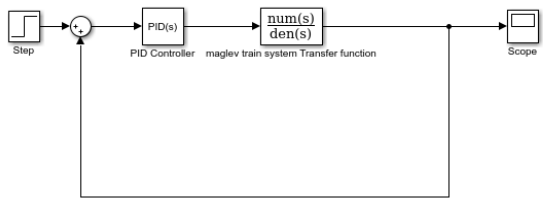

Fig 9 Simulink Diagram of Magnetic Levitation System using PID Controller

**4 Result and Discussion**
**4.1 Magnetic force versus current graph**
The magnetic force versus current graph of the Maglev train system is shown in Figure 10 below.

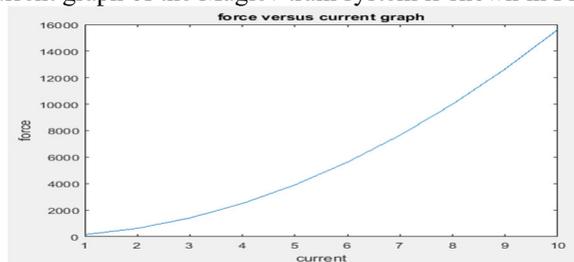

Fig 10. Magnetic force versus current graph plot

**4.2 Maglev train system simulation response**
The simulation output for Maglev train system without controller and Step Response of PID Auto-tuning for Maglev System is shown in Figure 11 and Figure 12 respectively.

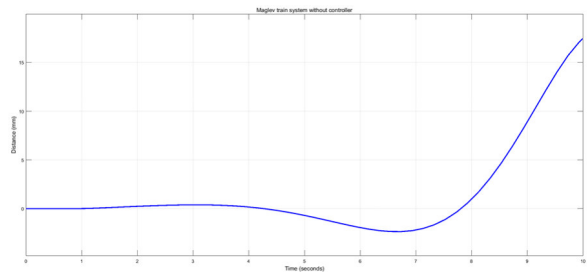

Fig 11 Maglev train system without controller

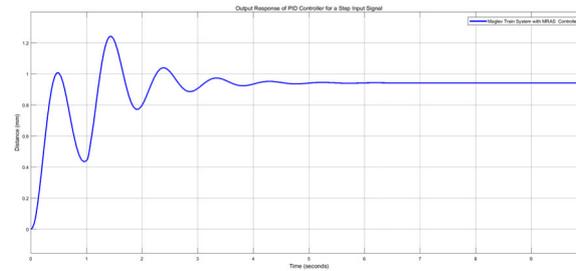

Fig 12 Step Response of PID Auto-tuning for Maglev System

### 4.3 Comparison of the Proposed Controllers
The output response of PID, FUZZY and MRAS Controllers for a step input is shown in Figure 14 below.

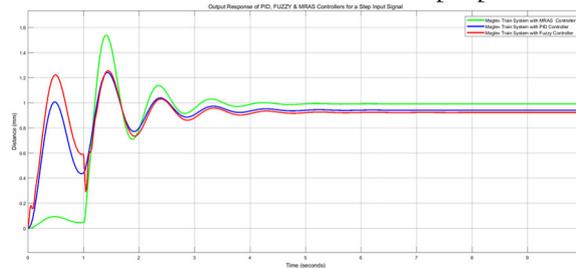

Fig 14. Output response of PID, FUZZY and MRAS Controllers for a step input.

The output response of maglev train system with different time period is shown in Figure 15 below.

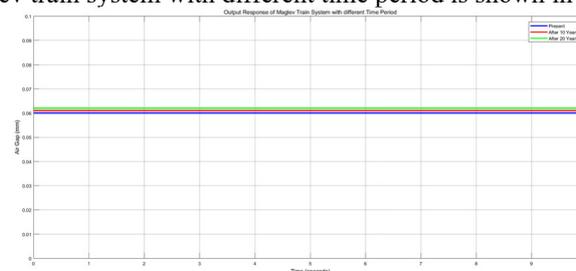

Fig 15. Output response of maglev train system with different time period

### 4.4 Numerical values of the Performance of PID, MRAS and Fuzzy Controllers
The numerical values of the proposed controllers is shown in Table 2 below.

Table 2. Numerical values of the proposed controllers

| Controller | Rise time (m sec) | Settling time (sec) | Percent Overshoot (%) | Steady state value |
|---|---|---|---|---|
| PID | 251.003 | 5.52 | 7.470 | 1 |
| MRAS | 150.897 | 5 | 55.469 | 0.93 |
| FUZZY | 264.604 | 5.9 | 32.854 | 0.89 |

The controller output response for different time period is shown in Table 3 below.

Table 3 Controller output response for different time period

| Controller output | | | | |
|---|---|---|---|---|
| Time | Max Overshoot | Rise time (sec) | Settling time (sec) | Percent Overshoot (%) |
| present | 0.0513 | 0.0523 | 0.9898 | 2.6 |
| After 10 years | 0.0518 | 0.0589 | 1.014 | 2.73 |
| After 20 years | 0.0525 | 0.0652 | 1.122 | 2.78 |

The Change of air gap with different time period is shown in Table 4 below.

Table 4. Change of air gap with different time period

| Period | Air gap (m) |
|---|---|
| Present | 0.06 |
| After 10 Years | 0.061 |
| After 20 Years | 0.062 |

**5 Conclusion**

Magnetic levitation system is inherently unstable system, because of the device nonlinearity. The output of the magnetic levitation device is determined and analyzed.

The simulation result showed that the settling time of MRAS controller is smaller than the settling time of PID and Fuzzy Controller. The rising time of MRAS controller is smaller than the rising time of PID and Fuzzy Controller. But the percentage overshoot of PID controller is very good when compared with Fuzzy controller and MRAS controller. And the controller can track the gap change and it could re-arrange itself with the gap change occur by change of time. Finally the simulation result prove the effectiveness of the MRAS controller.

**References**
[7]. Fabio C "High Speed Intercity and Urban Passenger Transport Maglev Train Technology Review: A Technical and Operational Assessment" Proceedings Series of ASME/IEEE Joint Rail Conference, 2019.
[8]. Zhai et al. "Calculation and Evaluation of Load Performance of Magnetic Levitation System in Medium Low Speed Maglev Train" International Journal of Applied Electromagnetics and Mechanics, Vol. 61 No. 4, pp. 519-536, 2019.
[9]. Wenk M et al. "Practical Investigation of Future Perspectives and Limitations of Maglev Technologies" Transportation Systems and Technology. Vol. 4, pp. 85-104, 2018.
[10]. Yang Y et al. "Study on the Optimization of Linear Induction Motor Traction System for Fast-Speed Maglev Train" Transportation Systems and Technology, Vol. 4, pp. 156-164, 2018.
[11]. Sonali Yadav et al. "A Review Paper on High Speed Maglev with Smart Platform Technology" International Journal for Scientific Research & Development, Vol. 6, Issue 09, 2018.
[12]. Yougang Sun et al. "Nonlinear Dynamic Modeling and Fuzzy Sliding Mode Controlling of Electromagnetic Levitation System of Low Speed Maglev Train" Journal of Vibro Engineering, Vol. 19, pp. 328-342, 2017.
[13]. Jae Hoon Jeong et al. "Analysis and Control of Electromagnetic Coupling Effect of Levitation and Guidance Systems for Semi High Speed Maglev Train Considering Current Direction" IEEE Transactions on Magnetics, vol. 53, Issue: 6, 2017.
[14]. Prachita Mame et al. "Magnetic Levitation Transportation System" International Journal of Students' Research in Technology & Management. Vol. 4, 2015.